\documentclass{osa-article}

\journal{osajournal}

\articletype{Research Article}
\begin{document}

\title{Cavity electro-optics in thin-film lithium niobate for efficient microwave-to-optical transduction}

\author{Jeffrey Holzgrafe,\authormark{1} Neil Sinclair,\authormark{1,2} Di Zhu,\authormark{1,3}  Amirhassan Shams-Ansari,\authormark{1} Marco Colangelo,\authormark{3} Yaowen Hu,\authormark{1,4} Mian Zhang,\authormark{1,5} Karl K. Berggren,\authormark{3} and Marko Lon\v{c}ar\authormark{1,*}}

\address{\authormark{1}John A. Paulson School of Engineering and Applied Sciences, Harvard University, 29 Oxford Street, Cambridge, Massachusetts 02138, USA\\
\authormark{2}Division of Physics, Mathematics and Astronomy, and Alliance for Quantum Technologies, California Institute of Technology, 1200 E. California Boulevard, Pasadena, California 91125, USA\\
\authormark{3}Research Laboratory of Electronics, Massachusetts Institute of Technology, 50 Vassar Street, Cambridge, Massachusetts 02139, USA\\
\authormark{4}Department of Physics, Harvard University, 17 Oxford Street, Cambridge, Massachusetts 02138, USA\\
\authormark{5}HyperLight Corporation, 501 Massachusetts Avenue, Cambridge, Massachusetts 02139, USA}

\email{\authormark{*}loncar@seas.harvard.edu} 



\begin{abstract*}
Linking superconducting quantum devices to optical fibers via microwave-optical quantum transducers may enable large scale quantum networks. For this application, transducers based on the Pockels electro-optic (EO) effect are promising for their direct conversion mechanism, high bandwidth, and potential for low-noise operation. However, previously demonstrated EO transducers require large optical pump power to overcome weak EO coupling and reach high efficiency. Here, we create an EO transducer in thin-film lithium niobate, leveraging the low optical loss and strong EO coupling in this platform. We demonstrate a transduction efficiency of up to $2.7\times10^{-5}$, and a pump-power normalized efficiency of $1.9\times10^{-6}/$\textmu W. The transduction efficiency can be improved by further reducing the microwave resonator's piezoelectric coupling to acoustic modes, increasing the optical resonator quality factor to previously demonstrated levels, and changing the electrode geometry for enhanced EO coupling. We expect that with further development, EO transducers in thin-film lithium niobate can achieve near-unity efficiency with low optical pump power.
\end{abstract*}

\section{Introduction}
Recent advances in superconducting quantum technology \cite{wendinQuantum2017} have created interest in connecting these devices and systems into larger networks. Such a network can be built from relatively simple quantum interconnects \cite{awschalomDevelopment2020} based on direct transmission of the few-photon microwave signals used in superconducting quantum devices \cite{kurpiersDeterministic2018}. However, the practical range of this approach is limited by the strong attenuation and thermal noise that microwave fields experience at room temperature. Therefore, quantum interconnects based on optical links have been explored as an alternative, because they provide long attenuation lengths, negligible thermal noise, and high bandwidth \cite{obrienPhotonic2009}. Connecting optical networks with superconducting quantum technologies requires the creation of a quantum transducer capable of converting single photons between microwave and optical frequencies \cite{lambertCoherent2019, laukPerspectives2019}. Such a transducer offers a promising route toward both large-scale distributed superconducting quantum networks and the scaling of superconductor quantum processors beyond single cryogenic environments \cite{barzanjehReversible2012}. Furthermore, single-photon microwave-to-optical transduction can be used to create high efficiency modulators \cite{gevorgyanTriply2020}, detectors for individual microwave photons \cite{bagciOptical2014}, and multiplexed readout of cryogenic electronics \cite{deceaReadout2020}.

An ideal quantum transducer performs a unitary transformation on microwave and optical modes. Practically, this level of performance implies efficiency approaching $100\%$, low noise, and enough bandwidth for the desired signals \cite{zeuthenFigures2017a}. Some of the most promising approaches toward quantum transduction have used electro- or piezo- optomechanical devices \cite{andrewsBidirectional2014, higginbothamHarnessing2018, midoloNanooptoelectromechanical2018, safavi-naeiniControlling2019, bochmannNanomechanical2013, fanAluminum2013, balramCoherent2016a, vainsencherBidirectional2016, jiangEfficient2020, forschMicrowavetooptics2020, shaoMicrowavetooptical2019, han10GHz2020}, nonlinear crystals that display a Pockels electro-optic (EO) effect \cite{ilchenkoWhisperinggallerymode2003, strekalovEfficient2009, ruedaEfficient2016, witmerHighQ2017, fanSuperconducting2018, witmerOnChip2019}, trapped atoms \cite{vogtEfficient2019, hanCoherent2018}, crystals doped with rare-earth ions \cite{fernandez-gonzalvoCoherent2015, bartholomewOnchip2019}, and optomagnonic devices \cite{hisatomiBidirectional2016}. Although quantum coherent performance has proved challenging to realize, the optomechanical approach --- the leading platform to date --- has achieved bidirectional operation \cite{andrewsBidirectional2014}, high efficiency \cite{higginbothamHarnessing2018, jiangEfficient2020}, and single-quantum scale noise levels \cite{forschMicrowavetooptics2020, mirhosseiniQuantum2020}. Optomechanical devices provide strong interactions with both microwave and optical fields, but their reliance on an intermediate mechanical mode in the transduction process creates several challenges for near-unitary operation. The low frequency (MHz) mechanical modes used in membrane electro-optomechanical transducers \cite{andrewsBidirectional2014, higginbothamHarnessing2018} result in strong thermal noise even at temperatures below $100\, \mathrm{mK}$, and make the transducers susceptible to low-frequency technical noise sources. Although piezo-optomechanical devices that use GHz frequency mechanical modes  \cite{midoloNanooptoelectromechanical2018, safavi-naeiniControlling2019, bochmannNanomechanical2013, fanAluminum2013, balramCoherent2016a, vainsencherBidirectional2016, jiangEfficient2020, forschMicrowavetooptics2020, shaoMicrowavetooptical2019, han10GHz2020} are not susceptible to these issues, such devices display optical-pump-induced heating of the mechanical resonator that adds to the transduction noise. This heating is difficult to avoid in piezo-optomechanical devices due to their colocalization of optical and mechanical modes, as well as their suspended and thermally isolated nature \cite{meenehanSilicon2014}. Pump-induced heating can be addressed using pulsed-pump schemes \cite{meenehanPulsed2015}, but this approach requires trade-offs between efficiency, noise, and repetition rate \cite{forschMicrowavetooptics2020, mirhosseiniQuantum2020}. 

The desire for lower noise, higher efficiency, and faster repetition rates has motivated research into cavity-based EO transducers \cite{tsangCavity2010, tsangCavity2011, javerzac-galyOnchip2016, ruedaEfficient2016, witmerHighQ2017, fanSuperconducting2018, witmerSiliconorganic2019, ilchenkoWhisperinggallerymode2003, strekalovEfficient2009}, in which microwave fields directly modulate light using an EO nonlinearity of the host material. This approach avoids any intermediate mechanical modes and may allow for lower noise owing to the strong thermal contact and spatial separation of microwave and optical resonators in these devices. Previous EO transducers have used bulk lithium niobate \cite{ilchenkoWhisperinggallerymode2003, strekalovEfficient2009, ruedaEfficient2016}, aluminum nitride \cite{fanSuperconducting2018} and hybrid silicon-organic \cite{witmerSiliconorganic2019} platforms. These devices have demonstrated bidirectional operation and on-chip efficiency as high as $2\%$ \cite{fanSuperconducting2018}, yet efficiencies remain low and would require large (${\sim} 1\, \mathrm{W}$) optical pump powers to reach near-unity efficiency. The efficiency of EO transducers can be improved by minimizing the loss rates of the resonators and enhancing the EO interaction strength. Toward this end, here we use a thin-film lithium niobate platform, which combines a large EO coefficient of $32\, \mathrm{pm/V}$, tight confinement of the optical mode to enable a strong EO coupling \cite{wangIntegrated2018}, and the ability to realize low-loss optical resonators with demonstrated quality factors (Q) of $10^7$ \cite{zhangMonolithic2017}.

Specifically, we describe an EO transducer made from a thin-film lithium niobate photonic molecule \cite{zhangElectronically2019, soltaniEfficient2017} integrated with a superconducting microwave resonator and demonstrate an on-chip transduction efficiency of greater than $10^{-6}/$\textmu W of optical pump power for continuous-wave signals. The triple-resonance photonic molecule design of our device maximizes transduction efficiency by ensuring that both the pump light and the upconverted optical signal are resonantly enhanced by low-loss optical modes. We also reduce undesired piezoelectric coupling in the microwave resonator by engineering bulk acoustic wave resonances in the device layers. Finally, we discuss the future potential for thin-film lithium niobate cavity electro-optic transducers and show that with straightforward improvements the efficiency can be increased to near unity for ${\sim} 100$ \textmu W of optical pump power.

\section{Device design and characterization}
\begin{figure}[h!]
\centering\includegraphics[width=7cm]{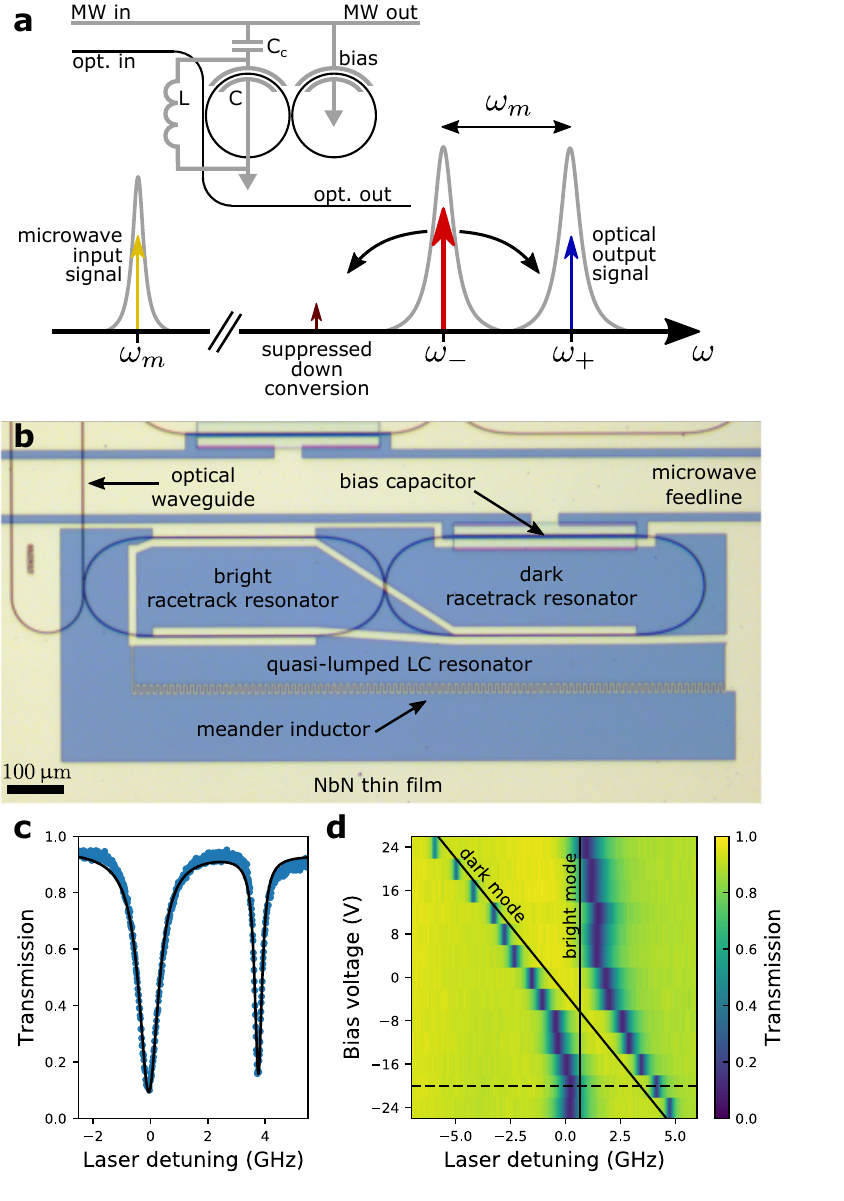}
\caption{A superconducting cavity electro-optic transducer on thin-film lithium niobate. (a) Top: Device schematic. Black and grey lines represent optical waveguides and superconducting wires, respectively. The coupled optical ring resonators are modulated by a microwave (MW) resonator (made from inductor L and capacitor C) that is capacitively coupled to the bus ($\mathrm{C_c}$). The optical resonance frequencies can be tuned by a bias capacitor. Bottom: Frequency-domain diagram of the transduction scheme. A pump laser is tuned into resonance with the red optical mode at $\omega_-$. Photons in the microwave resonator at $\omega_m$ can be upconverted to the blue optical mode at $\omega_+$ by sum-frequency generation. (b) Optical micrograph of the transducer, showing optical waveguides (black) and the niobium nitride superconducting film (light yellow). The capacitor of the quasi-lumped LC resonator modulates the optical racetrack resonators. A DC voltage on the microwave feedline controls the detuning between optical modes. (c) Optical transmission spectrum near $1586\, \mathrm{nm}$ of a pair of photonic-molecule optical modes with a $-20\, \mathrm{V}$ bias. (d) Optical transmission spectra at other bias voltages display an anticrossing between modes in the bright and dark racetrack resonators. Dashed line shows data in (c). \label{fig1}}
\end{figure}

The operating principle of our transducer is illustrated in Fig. \ref{fig1}(a). Two lithium niobate optical ring resonators are evanescently coupled to create a pair of hybrid photonic-molecule modes, with a strong optical pump signal tuned to the red optical mode at $\omega_-$. A superconducting microwave resonator with resonance frequency $\omega_m$ modulates the optical pump signal, upconverting photons from the microwave resonator to the blue optical mode at $\omega_+$.

This transduction process can be effectively described by a beamsplitter interaction Hamiltonian

\begin{equation}
    H_I = \hbar g_0 \sqrt{n_-} \left(b^\dag a_+ + b a_+^\dag\right),
\end{equation}
where $g_0$ is the single-photon EO interaction strength, $n_-$ is the number of (pump) photons in the red optical mode, while $b$ and $a_+$  are the annihilation operators for the microwave and blue optical modes, respectively. The interaction strength $g_0$ is determined by the microwave resonator's total capacitance, the overlap between microwave and optical modes, as well as the EO coefficient of the host material. We use a thin-film superconducting LC resonator and an integrated lithium niobate racetrack resonator \cite{wangIntegrated2018} to optimize $g_0$. The on-chip transduction efficiency $\eta$ for continuous-wave signals depends on both this interaction strength and the loss rates of the modes (see \textcolor{urlblue}{Supplement Section 1}),

\begin{equation}
    \eta = \frac{\kappa_{m,\mathrm{ex}}\kappa_{+,\mathrm{ex}}}{\kappa_{m} \kappa_+} \times \frac{4C}{\left(1+C\right)^2},
    \label{eta}
\end{equation}
where $\left(\kappa_{m,\mathrm{ex}}, \kappa_{m}\right)$ and $\left( \kappa_{+,\mathrm{ex}}, \kappa_+\right)$ are the external and total loss rates for the microwave and blue optical modes, respectively, and $C=\frac{4g_0^2n_-}{\kappa_m\kappa_+}$ is the cooperativity. The first term in Equation (\ref{eta}) represents the efficiency of a photon entering and exiting the transducer. To maximize this photon coupling efficiency, the resonators in our device are strongly overcoupled.

A microscope image of our device is shown in Fig. \ref{fig1}(b). Light is coupled from an optical fiber array onto the chip using grating couplers with ${\approx} 10\, \mathrm{dB}$ insertion loss. The photonic molecule optical modes are created using evanescently coupled racetrack resonators made from $1.2$ \textmu m-wide rib waveguides in thin-film lithium niobate atop a $4.7$ \textmu m-thick amorphous silicon dioxide layer on a silicon substrate. The optical waveguides are cladded with a $1.5$ \textmu m-thick layer of amorphous silicon dioxide. The fabrication process for these optical resonators is described in detail in Ref. \cite{zhangElectronically2019}. To create the superconducting resonator, a ${\approx}40\, \mathrm{nm}$-thick niobium nitride film is deposited on top of the cladding by DC magnetron sputtering \cite{daneBias2017} and patterned using photolithography followed by $\mathrm{CF_4}$ reactive ion etching. The detuning between the optical modes can be controlled using a bias capacitor on the dark (i.e. not directly coupled to the bus waveguide) racetrack resonator. 

The optical transmission spectrum displayed in Fig. \ref{fig1}c shows a typical pair of photonic-molecule optical modes. The internal loss rate for these optical modes are $\kappa_{\pm, \mathrm{in}}/2\pi \approx 2\pi \times 130 \, \mathrm{MHz}$, corresponding to an intrinsic quality factor of $1.4\times10^6$. As shown in Fig. \ref{fig1}d, we observe a clear anticrossing between bright and dark resonator modes when tuning the bias voltage and observe a minimum optical mode splitting of $3.1\,  \mathrm{GHz}$. For far-detuned optical modes, the bright resonator mode has a total loss rate $\kappa_{\mathrm{bright}}\approx 2\pi \times 1.0\,  \mathrm{GHz} $, indicating that the optical modes are strongly overcoupled to the bus waveguide.

\begin{figure}[h!]
\centering\includegraphics[width=7cm]{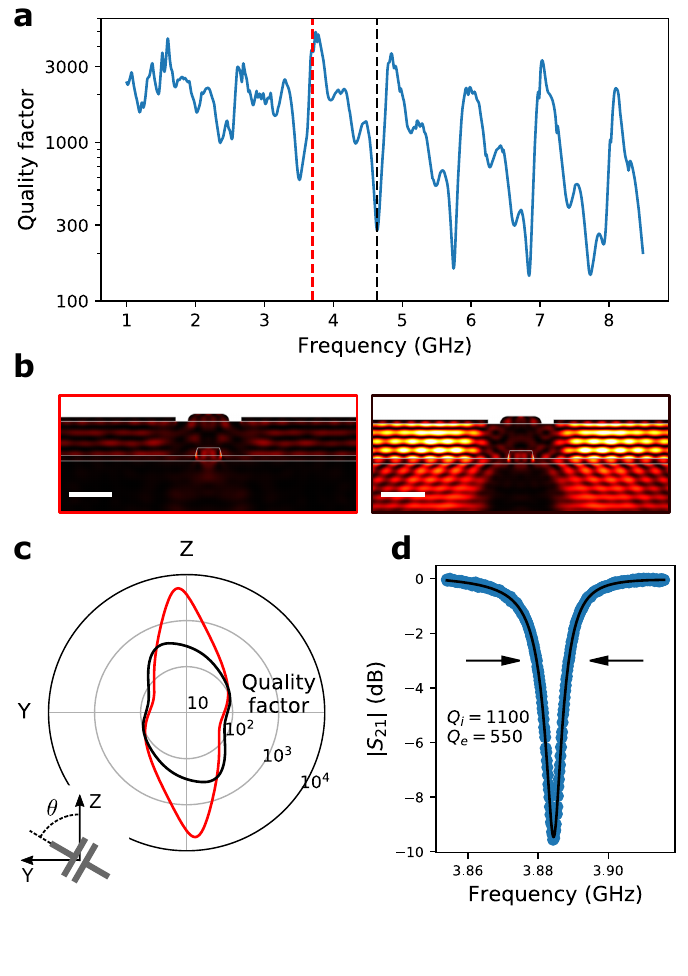}
\caption{Piezoelectric loss in lithium niobate. (a) Frequency dependence of the simulated microwave resonance loss caused by piezoelectric coupling to acoustic modes. Frequencies near bulk acoustic wave modes (e.g. black dashed line at $4.6\, \mathrm{GHz}$) display strong loss, but relatively low-loss performance can be achieved for frequencies far detuned from bulk acoustic modes (e.g. red dashed line at $3.7\, \mathrm{GHz}$). (b) Simulated acoustic energy density profiles of the device cross section for $3.7\, \mathrm{GHz}$ (red border) and $4.6\, \mathrm{GHz}$ (black border). White scale bars are $2\,$\textmu m long. (c) Simulated microwave quality factor for different orientations of the capacitor with respect to the crystal axes of lithium niobate, at $3.7\, \mathrm{GHz}$ (red) and $4.6\, \mathrm{GHz}$ (black). (d) The measured microwave transmission spectrum of the microwave resonator at a temperature $T=1\, \mathrm{K}$. \label{fig2}}
\end{figure}

Lithium niobate has a strong piezoelectric susceptibility, which gives the microwave resonator a loss channel to traveling acoustic modes \cite{scigliuzzoPhononic2019}. To investigate this loss mechanism, we perform a two-dimensional simulation of a cross section of the waveguide and resonator capacitor (see \textcolor{urlblue}{Supplement Section 2}). The simulated intrinsic microwave quality factor due to piezoelectric loss displays a strong frequency dependence, as shown in Fig. \ref{fig2}(a). This frequency dependence is caused by low quality-factor bulk acoustic modes --- illustrated in Fig. \ref{fig2}(b) --- that form in the thin-film layers of our device, which resonantly enhance the coupling between the microwave resonator and acoustic fields. Lower loss can be achieved by designing the microwave resonance frequency to avoid the bulk acoustic resonances. The relative orientation of the capacitor and the lithium niobate crystal axes also strongly affects the microwave loss. Figure \ref{fig2}(c) shows that the intrinsic microwave quality factor is maximized when the electric field produced by the capacitor is oriented close to the Z-axis of the lithium niobate crystal, which is also the condition that maximizes electro-optic response. Using these considerations, we designed a microwave resonator which has a measured intrinsic quality factor higher than $10^3$ at a temperature of $1\, \mathrm{K}$, as shown in Fig. \ref{fig2}(d). 

\section{Microwave to optical transduction}
\begin{figure}[h!]
\centering\includegraphics[width=8cm]{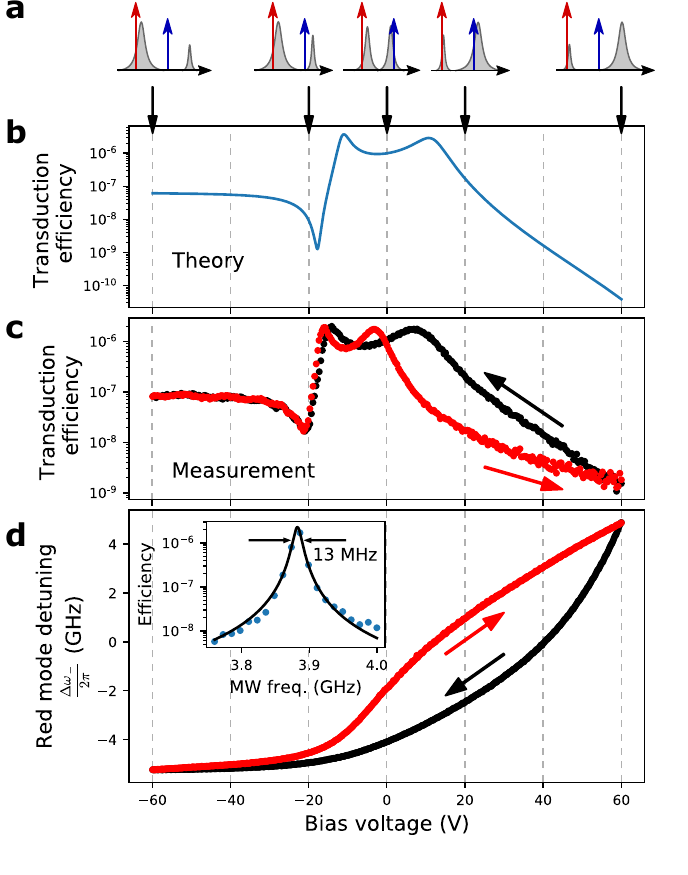}
\caption{Detuning dependence of microwave-to-optical photon transduction. (a) Frequency-domain illustrations of the optical modes for the bias voltages indicated by the downward-facing arrows. (b) Predicted dependence of the transduction efficiency on the bias voltage. This prediction is based on a model that includes transduction into both the red and blue optical modes, and independently measured and calculated device parameters. (c) Measured dependence of the transduction efficiency on the bias voltage. Red and black traces correspond to sweeps of increasing and decreasing voltage, respectively. (d) The frequency detuning of the red optical mode caused by the bias voltage during the transduction experiment. Inset: The dependence of the highest transduction efficiency on the frequency of the microwave drive displays a wide bandwidth of $13\, \mathrm{MHz}$. All data shown for $-30 \,\mathrm{dBm}$ on-chip optical pump power. \label{fig3}}
\end{figure}

To measure the transduction efficiency of our device, we locked the frequency of the pump to be near-resonant with the red optical mode (side-of-fringe locking) and sent a resonant microwave signal into the device. The pump and upconverted optical signal were collected and sent to an amplified photodetector, which produced a beat note at the input microwave frequency. We inferred the transduction efficiency from this beat note by calibrating the input optical power, system losses and detector efficiency (see \textcolor{urlblue}{Supplement Section 3}). During this transduction efficiency measurement, we swept the bias voltage in a triangle waveform with a period of ${\approx} 1\, \mathrm{min}$ to vary the splitting between the optical modes. The pump light remained locked to the red optical mode throughout the measurement. Figure \ref{fig3}(a) illustrates the optical modes and signals over the course of the bias voltage sweep.

The results of this measurement are depicted in Fig. \ref{fig3}. The two maxima in transduction efficiency near $\pm10\, \mathrm{V}$ (Figs. \ref{fig3}(b) and (c)) correspond to the two cases in which a triple-resonance condition is met and the upconverted light is resonant with the blue optical mode. For large negative bias voltages, the blue mode is far-detuned, and most of the upconverted light is generated in the red optical mode by a double-resonance process involving just the red optical mode and the microwave mode (see \textcolor{urlblue}{Supplement Section 1.3}). This process does not depend on the resonance frequency of the blue optical mode, so the transduction efficiency is nearly independent of the bias voltage in this regime. Destructive interference of upconverted light produced in the red and blue optical modes (created by the double- and triple-resonance processes, respectively) causes the transduction efficiency minimum near $-20 \mathrm{V}$. For large positive bias voltage, the red optical mode is undercoupled and has a narrow linewidth, so the double-resonance transduction process is weak. The measured data presented in Fig. \ref{fig3}(c) shows good correspondence to our analytical model (see \textcolor{urlblue}{Supplement Section 1}) based on independently measured and estimated device parameters, shown in Fig. \ref{fig3}(b).

The measured transduction efficiency features strong hysteresis when varying the bias voltage, which is caused by hysteresis in the detuning of the optical modes, shown in Fig. \ref{fig3}(d). We observed that this hysteresis could be reduced by lowering the optical pump power and sweeping the voltage bias faster. Based on the slow timescale (seconds for $-30\, \mathrm{dBm}$ on-chip optical pump power), we attribute the hysteresis to photoconductive and photorefractive effects in lithium niobate \cite{jiangFast2017, gunterPhotorefractive2006}. These effects are caused by optical excitation of charge carriers in the lithium niobate waveguide, which can migrate to create built-in electric fields that shift the optical resonance frequencies through the EO effect.

\begin{figure}[h!]
\centering\includegraphics[width=8cm]{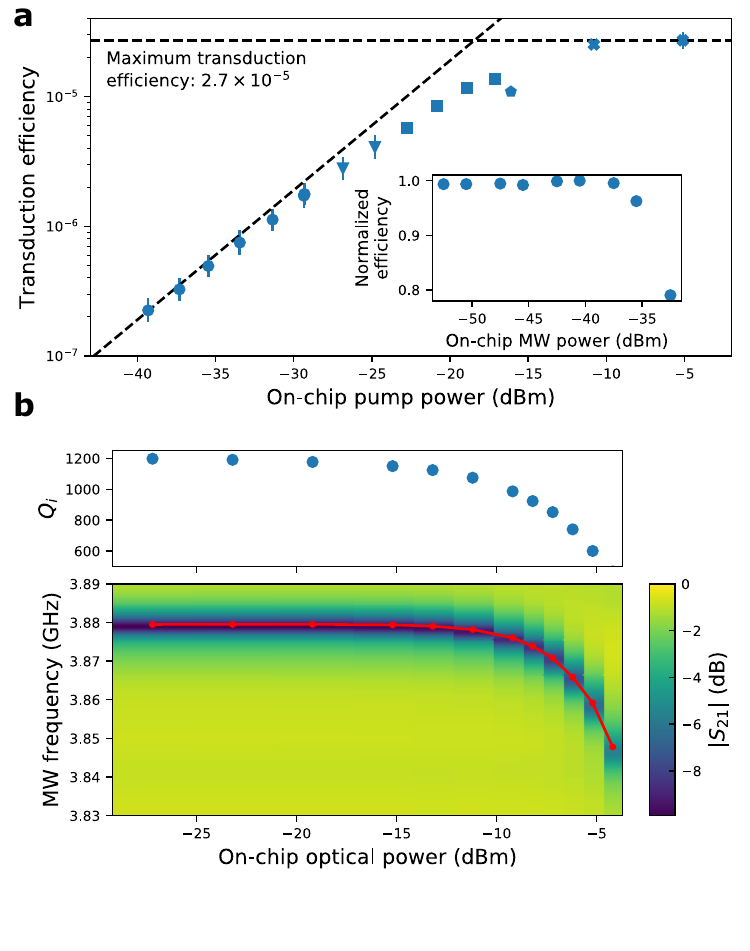}
\caption{Power dependence of microwave-to-optical photon transduction. (a) The highest transduction efficiency measured during a sweep of the bias voltage for different optical pump powers. The different markers (circles, triangles, etc.) correspond to sets of measurements performed on different optical modes - see main text for details. Inset: The dependence of the transduction efficiency on microwave input power for $-30\, \mathrm{dBm}$ on-chip optical pump power. (b) Degradation of the microwave resonator due to absorption of pump light. The plots show the optical power dependence of the microwave transmission spectrum (bottom, red line denotes resonance frequency) and the inferred intrinsic microwave quality factor (top). \label{fig4}}
\end{figure}

We measured the bandwidth of the transducer by varying the frequency of the input microwave drive and measuring the highest transduction efficiency reached during a bias voltage sweep. The inset of Fig. \ref{fig3}(d) shows that our transducer has a $3\, \mathrm{dB}$ bandwidth of $13\, \mathrm{MHz}$, slightly larger than the measured $10\, \mathrm{MHz}$ linewidth of the microwave resonator. This discrepancy is caused by the nonlinear response of the NbN microwave resonator for high microwave power, which leads to an apparent resonance broadening \cite{abdoNonlinear2006} (see \textcolor{urlblue}{Supplement Section 4}). To reduce measurement noise, here we use a relatively large microwave power ($-38\, \mathrm{dBm}$ on-chip), which causes a small degree of nonlinear broadening. This nonlinearity also leads to reduced transduction efficiency for large input microwave powers (Fig. \ref{fig4}(a) inset).

Figure \ref{fig4}(a) shows the highest transduction efficiency measured during a bias voltage sweep for different optical pump powers. During this measurement, we found that the detuning between the optical modes could change by several $\mathrm{GHz}$ when the optical pump power was varied, likely due to the same effects that cause the hysteresis described earlier. To measure the transduction efficiency at the triple-resonance point for each power level, we performed measurements on several pairs of photonic-molecule modes. For the highest optical pump powers used in this study (denoted by crosses in Fig. \ref{fig4}(a)), we modulated the optical pump to extinction at a rate of $20\, \mathrm{kHz}$ and with a $10\%$ duty cycle. The lower average power in these modulated-pump measurements resulted in smaller power-dependent detunings and more stable resonances. 

In the low-power regime ($<-30 \, \mathrm{dBm}$) we observe that the transduction efficiency scales linearly with pump power at a rate of $(1.9 \pm 0.4)\times 10^{-6}/$\textmu W. From this and the measured loss rates of the resonators, we estimate the single-photon coupling rate of our transducer to be $g_0 = 2\pi \times 650 \pm 70\, \mathrm{Hz}$, comparable in magnitude to the predicted $g_0 = 2\pi \times 830\, \mathrm{Hz}$ (see \textcolor{urlblue}{Supplement Section 1.4}), yet slightly lower than expected. This difference is likely due to variations in the as-fabricated geometry of the device. 

The transduction efficiency begins to saturate at $(2.7\pm 0.3)\times10^{-5}$, the highest-measured efficiency for this transducer. This saturation is caused by optical absorption in the microwave resonator, which generates quasiparticles that shift the resonance frequency and increase the loss rate of the resonator\cite{zmuidzinasSuperconducting2012}. Figure \ref{fig4}(b) shows the optical-power dependence of the microwave resonator's properties. We find that the quasiparticle-induced changes in the microwave resonator are independent of whether the pump laser is tuned on- or off-resonance with an optical mode, which suggests that the absorbed light does not come from the optical resonator itself. Instead, light scattered at the fiber array and grating couplers is likely the dominant contribution to the quasiparticle loss.

\section{Discussion and Conclusion}
The transduction efficiency demonstrated here falls well below the requirements for a useful quantum transducer. However, several straightforward improvements can be made to the transducer to increase this figure of merit (see \textcolor{urlblue}{Supplement Section 5}). First, optical quality factors above $10^7$ have been demonstrated in thin-film lithium niobate \cite{zhangMonolithic2017}, suggesting that the optical loss rates seen here can be reduced by roughly 10-fold, leading to a 100-fold improvement in transduction efficiency. Second, the microwave resonator loss rate can be reduced through improved engineering of the bulk acoustic waves to which the microwave resonator couples. For example, simulations suggest that suspending the lithium niobate layer can reduce the microwave loss by more than 10-fold. The quasiparticle losses caused by stray light absorption for higher optical pump powers can be made negligible by changing the design of the sample mount and optical fiber coupling. By using these and other (see \textcolor{urlblue}{Supplement Section 5}) interventions, we predict that near-unity transduction efficiency can be achieved for optical pump powers of ${\sim} 100$ \textmu W. Although we did not demonstrate optical-to-microwave transduction (only microwave-to-optical) due to a low signal-to-noise ratio, we note that the transduction process described here is fully bidirectional \cite{fanSuperconducting2018}.

In this work, we have demonstrated transduction between microwave and optical frequencies using a thin-film lithium niobate device. The photonic molecule design of our transducer enables straightforward tuning of the optical modes using a bias voltage, ensures strong suppression of the downconverted light that acts as a noise source, and takes full advantage of the large electro-optic coefficient in lithium niobate. We have described how the piezoelectric coupling of the microwave resonator to traveling acoustic waves can be engineered to minimize loss in the microwave resonator. The advantages of an EO transducer - namely the system simplicity and the possibility for low-noise operation - and the opportunities for improved transduction efficiency suggest that further development of thin-film lithium niobate cavity electro-optics is warranted. 

During the preparation of this manuscript we became aware of a similar lithium niobate quantum transducer device reported by McKenna et al. \cite{mckennaCryogenic2020}.

\section*{Funding}
National Science Foundation (NSF) (ECCS-1839197, ECCS-1541959); Natural Sciences and Engineering Research Council of Canada (NSERC); AQT Intelligent Quantum Networks and Technologies (INQNET) research program; DOE/HEP QuantISED program grant, QCCFP (Quantum Communication Channels for Fundamental Physics), award number DE-SC0019219.

\section*{Acknowledgments}
This work was performed in part at the Center for Nanoscale Systems (CNS), a member of the National Nanotechnology Coordinated Infrastructure Network (NNCI), which is supported by the National Science Foundation under NSF award no. 1541959. CNS is part of Harvard University. D.Z. is supported by the Harvard Quantum Initiative (HQI) postdoctoral fellowship. The authors thank C.J. Xin, E. Puma, B. Machielse, C. Wang, Y. Qiu, and W. Oliver for helpful discussions and assistance with device fabrication.

\bigskip

\noindent See Supplement 1 for supporting content.

\bibliography{bib}

\pagebreak
\setcounter{equation}{0}
\setcounter{section}{0}
\setcounter{figure}{0}
\setcounter{table}{0}
\setcounter{page}{1}
\renewcommand{\theequation}{S\arabic{equation}}
\renewcommand{\thefigure}{S\arabic{figure}}
\renewcommand{\thetable}{S\arabic{table}}

\begin{center}
	\textbf{\large Supplement 1}
\end{center}

\section{Theory of cavity electro-optics for microwave-to-optical transduction}
\subsection{Optical modes}

The optical modes used in our device are delocalized between two evanescently coupled ring resonators. These optical modes are governed by the Hamiltonian 
\begin{equation}
    H_{\mathrm{opt}} = (\omega_o +\delta) a_1^\dag a_1 + (\omega_o-\delta) a_2^\dag a_2 + \mu \left( a_1^\dagger a_2 + a_1 a_2^\dagger \right),
\end{equation}
where $a_1$ and $a_2$ are respectively the annihilation operators for the bright and dark ring resonator modes, $2\delta$ is the detuning between ring resonator modes, and $\mu$ is the evanescent coupling rate. This Hamiltonian can be diagonalized by the Bologliubov transformation
\begin{align}
    a_+ &=  va_2 - ua_1, \nonumber\\
    a_- &= ua_2 + va_1.\label{bogo}
\end{align}

Note that $a_+$ and $a_-$ must obey a bosonic commutation relation $\left[a_i, a_j\right]=\delta_{ij}$, which requires that $u^2+v^2=1$. For convenience, we set $u=\cos{\frac{\theta}{2}}$ and $v=\sin{\frac{\theta}{2}}$, where $\theta$ is a hybridization parameter. The Hamiltonian will be diagonalized for $\tan{\theta} = \frac{\mu}{\delta}$, giving

\begin{equation}
     H_{\mathrm{opt}} = \omega_+ a_+^\dag a_+ + \omega_- a_-^\dag a_-,
\end{equation}
where the resonance frequencies are $\omega_{\pm} = \omega_0 \pm \sqrt{{\delta\omega}^2 + \mu^2}$. These optical system eigenmodes at frequency $\omega_{\pm}$ will be used to calculate the interaction Hamiltonian.

The loss rates of the photonic molecule modes change with the hybridization parameter $\theta$. In the resolved sideband approximation  ($2\mu \gg \kappa$, where $\kappa$ is the typical optical mode loss rate), Eq. \ref{bogo} also diagonalizes the open system, and the internal (i) and external (e) loss rates for the hybrid modes, $\kappa_{\pm,\mathrm{\{i,e\}}}$, are given by
\begin{align}
    \kappa_{+,\mathrm{\{i,e\}}} &= v^2\kappa_{2,\mathrm{\{i,e\}}} + u^2 \kappa_{1,\mathrm{\{i,e\}}}, \nonumber\\
    \kappa_{-,\mathrm{\{i,e\}}} &= u^2\kappa_{2,\mathrm{\{i,e\}}} + v^2 \kappa_{1,\mathrm{\{i,e\}}},
\end{align}
where $\kappa_{\{1,2\},\mathrm{\{i,e\}}}$ are the intrinsic and extrinsic loss rates for the bright and dark ring resonator modes.
\subsection{Triple-resonance transduction}

Our electro-optic transducer has three resonant modes. The two optical photonic-molecule modes have electric fields $\Vec{E}_{\pm}(\Vec{r}) = \sqrt{\frac{\hbar \omega_\pm}{2\epsilon (\Vec{r})V_{\pm}}} \left(\vec{\psi}_\pm(\Vec{r})a_\pm e^{-i\omega_\pm t} + \mathrm{h.c.} \right)$. The microwave mode has electric field $\Vec{E}_{m}(\Vec{r})= \sqrt{\frac{\hbar \omega_m}{2C d_\mathrm{eff}^2}} \left(\vec{\psi}_m(\Vec{r})b e^{-i\omega_m t} + \mathrm{h.c.} \right)$. Here $\omega_{\{\pm, m\}}$ are the resonance frequencies, $\epsilon$ is the dielectric permittivity, $V_\pm$ are effective mode volumes,  $a_\pm$ and $b$ are annihilation operators for the optical and microwave modes,  $\vec{\psi}_{\{\pm, m\}}$ are the field profiles\footnote{We assume a lumped-element model for the microwave mode and normalized it so  $|\psi_m|^2 =1$ at the center of the optical mode cross section in areas covered by the capacitor.}, $C$ is the total capacitance of the microwave resonator, and $d_\mathrm{eff}$ is a constant with dimensions of length that relates the voltage on the microwave resonator's capacitor to the electric field at the center of the optical waveguide. For our device geometry, all three modes are polarized approximately along the Z-axis of the nonlinear crystal in the region of interest, so we can use a scalar interaction approximation. The three-wave mixing process used in our device can be described by a nonlinear energy density \cite{armstrongInteractions1962}

\begin{equation}
    U = 2 \epsilon_0 \chi^{(2)} E_+ E_- E_m.
    \label{ABDP}
\end{equation}

The interaction Hamiltonian for this process can be obtained from Eq. \ref{ABDP} by inserting our expressions for the mode fields and considering only terms that vary slowly near the triple-resonance condition $\omega_m = \omega_+ - \omega_-$, which yields:
\begin{equation}
    H_I = \hbar g_0 \left(ba_- a_+^\dag + b^\dag a_-^\dag a_+ \right),
\end{equation}
The single-photon interaction strength is
\begin{equation}
    g_0 = \frac{\chi^{(2)} \omega_{\mathrm{opt}}}{d_{\mathrm{eff}}V_{\mathrm{opt}}n^2}\sqrt{\frac{\hbar \omega_m}{2C}} \int dV \psi_+^*\psi_-\psi_m,
    \label{g_0}
\end{equation}
where $n$ is the optical refractive index and the integral is taken over the nonlinear material. If a strong pump laser is tuned to the red optical mode, we can replace the $a_-$ operator with its classical expectation value $a_- \to \langle a_-\rangle = \sqrt{n_-} = \sqrt{\frac{\kappa_{-, \mathrm{e}}}{\Delta_-^2 + (\kappa_-/2)^2}}\sqrt{\frac{P}{\hbar\omega_l}}$,  where $\kappa_{-}=\kappa_{-, \mathrm{i}}+\kappa_{-,\mathrm{e}}$ is the total loss rates of the red optical mode, $\omega_l$ is the pump laser frequency, $\Delta_- = \omega_l-\omega_-$ is the pump detuning, and $P$ is the pump power. Moving to a frame where the optical modes rotate at the pump laser frequency, the full Hamiltonian for our system is 
\begin{equation}
    H = -\Delta_+ a_+^\dag a_+ + \omega_m b^\dag b + g \left(b^\dag a_+ + b a_+^\dag \right),
\end{equation}
where $\Delta_+ = \omega_l - \omega_+$ is the detuning of the pump from the blue optical mode, and $g=g_0\sqrt{n_-}$ is the pump-enhanced coupling rate.

We now use the above Hamiltonian to estimate the bidirectional transduction efficiency between continuous-wave (CW) optical and microwave signals. Consider two signal fields incident on the transducer: an optical signal $a_{\mathrm{in}}$ detuned from the pump by $\omega_p$, and a microwave signal $b_{\mathrm{in}}$ with frequency $\omega_{b\mathrm{in}}$. The semi-classical Heisenberg-Langevin equations of motion governing the interaction between the microwave and blue optical modes are
\begin{align}
    &\frac{d a_+}{d t} = -\left(-i\Delta_+ + \frac{\kappa_+}{2}\right) a_+ -igb+\sqrt{\kappa_{+,\mathrm{e}}}a_{\mathrm{in}}e^{-i\omega_p t}, \nonumber \\
    &\frac{d b}{d t} =-\left(i\omega_m + \frac{\kappa_m}{2}\right) b -ig a_+ +\sqrt{\kappa_{m,\mathrm{e}}}b_{\mathrm{in}}e^{-i\omega_{b\mathrm{in}} t},
    \label{heisenberg_langevin}
\end{align}
where $\kappa_{m}$ and $\kappa_{m, \mathrm{e}}$ are the total and external coupling rates for the microwave mode. The resonator modes couple to propagating output fields $a_{\mathrm{out}}$ and $b_{\mathrm{out}}$  via the input-output relations
\begin{align}
    &a_{\mathrm{out}} = a_{\mathrm{in}} - \sqrt{\kappa_{+,\mathrm{e}}}a_+,
    &b_{\mathrm{out}} = b_{\mathrm{in}} - \sqrt{\kappa_{b,\mathrm{e}}}b.
    \label{io_relations}
\end{align}
In the steady state, Eqs. \ref{heisenberg_langevin} and \ref{io_relations} yield the frequency-domain transduction scattering matrix
\begin{equation}
    \label{SMatrix}
    \begin{bmatrix}
    a_{\mathrm{out}}\\
    b_{\mathrm{out}}
    \end{bmatrix}
    =
    \begin{bmatrix}
    S_{\mathrm{oo}}&
    S_{\mathrm{oe}}\\
    S_{\mathrm{eo}}&
    S_{\mathrm{ee}}
    \end{bmatrix}
    \begin{bmatrix}
    a_{\mathrm{in}}\\
    b_{\mathrm{in}}
    \end{bmatrix},\\
\end{equation}
where
\begin{align*}
    &S_{\mathrm{oo}} = 1-\frac{\kappa_{+,\mathrm{e}}}{-i\left(\Delta_++\omega_p\right) +\frac{\kappa_+}{2}+\frac{g^2}{i\left(\omega_m-\omega_p\right)+\kappa_m/2}}, \\
    &S_{\mathrm{oe}} =     \frac{ig\sqrt{\kappa_{+,\mathrm{e}}\kappa_{b,\mathrm{e}}}}{\left[i\left(\omega_m-\omega_{b\mathrm{in}}\right) +\frac{\kappa_m}{2}\right]\left[-i\left(\Delta_++\omega_{b\mathrm{in}}\right)+\frac{\kappa_+}{2}\right]+g^2}, \\
    &S_{\mathrm{eo}} = \frac{ig\sqrt{\kappa_{+,\mathrm{e}}\kappa_{b,\mathrm{e}}}}{\left[i\left(\omega_m-\omega_p\right) +\frac{\kappa_m}{2}\right]\left[-i\left(\Delta_++\omega_p\right)+\frac{\kappa_+}{2}\right]+g^2}, \\
    &S_{\mathrm{ee}} = 1-\frac{\kappa_{m,\mathrm{e}}}{i\left(\omega_m-\omega_{b\mathrm{in}}\right)+\frac{\kappa_m}{2}+\frac{g^2}{-i\left(\Delta_++\omega_{b\mathrm{in}}\right)+\kappa_+/2}}.
\end{align*}

The conversion is symmetric, and the on-chip transduction efficiency is

\begin{equation}
\begin{split}
    \eta_{\{\mathrm{oe, eo}\}}(\omega) &= \lvert S_{\{\mathrm{oe, eo}\}} \rvert^2 ,
\end{split}
\end{equation}
where $\omega$ is the excitation frequency, and $C=\frac{4g^2n_-}{\kappa_+\kappa_m}$ is the electro-optic cooperativity. At the triple-resonance condition, where $\omega=-\Delta_+=\omega_m$, this efficiency takes the form
\begin{equation}
    \eta_{\{\mathrm{oe, eo}\}}(\omega) = \underbrace{\frac{\kappa_{m,e}\kappa_{+,e}}{\kappa_m \kappa_+}}_{\mathrm{extraction\ efficiency}}\times\underbrace{ \frac{4C}{\left(1+C\right)^2}}_{\mathrm{internal\ efficiency}}.
\end{equation}
The first term represents the efficiency associated with getting a photon into and out of the converter, while the second term gives the transduction efficiency inside the converter.

\subsection{Double-resonance transduction}
When the converter is operated far-detuned from the triple-resonance condition, a double-resonance transduction process can become a significant contribution to the total transduction efficiency. In this process, optical photons in the red optical mode can be scattered between the pump frequency and a blue-shifted sideband. This sideband field is far-detuned from the red optical mode relative the mode's linewidth in our experiments, so the transduction efficiency for this process is low, but it can be larger than that of the triple-resonance process when the splitting between red and blue optical modes is much larger than the microwave frequency. This double-resonance process is the origin of the bias-voltage independent response for large negative voltages in Figs. 3(b) and 3(c) of the main text. The nonlinear energy density that describes this double-resonance process is \cite{armstrongInteractions1962}
\begin{equation}
    U = \epsilon_0\chi^{(2)}E_mE_-^2
\end{equation}
which produces the Hamiltonian
\begin{equation}
    H_I =  \hbar g_{0,\mathrm{dr}} \left(ba_-^\dag a_- + b^\dag a_-^\dag a_- \right),
\end{equation}
where

\begin{equation}
    g_{0,\mathrm{dr}} = \frac{\chi^{(2)} \omega_{\mathrm{opt}}}{d_{\mathrm{eff}}V_{\mathrm{opt}}n^2}\sqrt{\frac{\hbar \omega_m}{2C}} \int dV \lvert \psi_+ \rvert^2\psi_m.
\end{equation}

Following the usual linearization procedure for the strongly pumped $a_-$ mode \cite{aspelmeyerCavity2014}, we approximate $a_- \approx \langle a_- \rangle + \delta a_-$, where $\delta a_-$ is a small fluctuating perturbation to the field in the red optical mode. Keeping terms of order $\langle a_- \rangle$, the linearized interaction Hamiltonian is
\begin{equation}
    H_I = \hbar g_{\mathrm{dr}} \left(\delta a_-^\dag + \delta a_-\right) \left( b^\dag + b \right),
\end{equation}
where $g_{\mathrm{dr}} = g_{0,\mathrm{dr}}\langle a_- \rangle$. This Hamiltonian contains both the desired beam-splitter terms and parametric amplification terms which cause optical down conversion, and since the pump is nearly resonant with the red optical mode, both types of terms are significant. In a frame where the optical mode rotates along with the laser, the semi-classical Heisenberg-Langevin equations of motion for double-resonance microwave-to-optical transduction are
\begin{align}
    &\frac{d \left(\delta a_-\right)}{d t} = -\left(-i\Delta_- + \frac{\kappa_-}{2}\right) \delta a_-  -ig_{\mathrm{dr}}\left( b + b^\dag \right), \nonumber \\
    &\frac{d b}{d t} =-\left(i\omega_m + \frac{\kappa_m}{2}\right) b -i g_{\mathrm{dr}} \left(\delta a_- + \delta a_-^\dag\right) +\sqrt{\kappa_{m,\mathrm{e}}}b_{\mathrm{in}}e^{-i\omega t}.
    \label{dr_heisenberg_langevin}
\end{align}
For simplicity, we assume that the double resonance process operates in the weak coupling regime, so that back-action of the optical fields on the microwave field can be neglected\footnote{i.e. the term $-i g_{\mathrm{dr}} \left(\delta a_- + \delta a_-^\dag\right)$ can be dropped}. We take the ansatz solution
\begin{align}
    \delta a_-(t) &= A_+ e^{-i\omega t} + A_- e^{i\omega t},\nonumber\\
    b(t) &= B e^{-i\omega t},
\end{align}
and find
\begin{align}
    B&= \frac{\sqrt{\kappa_{m,e}}b_{\mathrm{in}}}{i\left(\omega_m-\omega\right) + \kappa_m/2}\nonumber\\
    A_\pm&= \frac{-ig_{\mathrm{dr}}B}{-i\left(\Delta_-\pm\omega\right)+\kappa_-/2}.
\end{align}
The transmitted optical sideband field due to double-resonance transduction is $\delta a_{\mathrm{out}} = -\sqrt{\kappa_{-,e}} \delta a_-$, and hence the total apparent transduction efficiency\footnote{The existence of multiple optical sidebands in regimes where double-resonance transduction is significant means that transduction efficiency must be carefully defined. In our experiments, we measure only the transmission of a microwave signal from the transducer's input to the photoreceiver's output, and we cannot differentiate multiple optical sidebands. As such, we define transduction efficiency for multiple sidebands as the apparent transduction efficiency: i.e. the equivalent single-sideband transduction efficiency which would produce the observed signal. Note that this distinction between apparent and true transduction efficiency is significant only in far-detuned regimes of bias voltage sweeps, not near the triple-resonance condition where maximum transduction efficiency occurs.}, including both double- and triple- resonance transduction, is 
\begin{equation}
    \eta_{\mathrm{oe}} = \lvert S_{\mathrm{eo}} -\sqrt{\kappa_{-,e}}A_+ -\sqrt{\kappa_{-,e}}A_-^*  \rvert^2.
\end{equation}

\subsection{Estimating the electro-optic interaction strength}

The triple-resonance interaction strength $g_0$ (Eq. \ref{g_0}) can be cast in a form more useful for designing the transducer. Assuming that the electric field created by the capacitor is  oriented along the lithium niobate's Z crystal axis and uniform across the optical mode (a good approximation for our device geometry), and that the microwave resonator drives the optical resonators with opposite phase and behaves as a lumped-element system, we find
\begin{equation}
    g_0(\theta) = \frac{r_{33}n_e^2\hbar\omega_o \Gamma \alpha \sin{(\theta)}}{4d_{\mathrm{eff}}}\sqrt{\frac{\hbar\omega_m}{2C}},
\end{equation}
here $r_{33}= 2\chi^{(2)}/n^4$ is the relevant electro-optic coefficient, $n_e$ is the extraordinary refractive index of lithium niobate, $\Gamma$ is an optical mode confinement factor, $\alpha$ is an electrode coverage parameter with a maximum value of $2$ for full coverage of both optical resonators, and $\theta$ is the optical mode mixing parameter described above\footnote{Note that the values of $g_0$ used in the main text are quoted for the optical mode splitting (and hence the value of the hybridization parameter $\theta)$ which maximizes transduction efficiency. The maximum transduction efficiency is obtained for $\theta\approx 0.7\pi$ in our device.}. From this equation, it is clear that the interaction strength can be maximized by creating a microwave resonator with closely spaced electrodes, low total capacitance, and full coverage of the optical resonators. 

The calculated values for key device parameters are given in Table \ref{tab:dev_params}. Using these results, we estimate $g_0(\theta=\pi)=2\pi \times 1.0\ \mathrm{kHz}$.

\begin{table}[htbp]
\centering
\caption{\bf Transducer device parameters for calculating the electro-optic interaction strength}
\begin{tabular}{ccc}
\hline
Parameter & Value & Estimation method \\
\hline
$\Gamma$ & $0.93$ & Optical mode simulation  \\
$\alpha$ & 0.72 & Calculated from device geometry \\
$d_{\mathrm{eff}}$ & $15$\,\textmu m & EO cross section simulation \\
$C$ & $120 \, \mathrm{fF}$ & Full-wave simulation as in \cite{finkQuantum2016}\\

\hline
\end{tabular}
  \label{tab:dev_params}
\end{table}

\section{Simulating piezoelectric loss}
We use a two-dimensional finite element model to simulate the piezoelectric loss of the microwave resonator. In this frequency-domain simulation, a voltage is applied to the capacitor electrodes at frequency $\omega$, and the time-averaged electrostatic energy $E_{\mathrm{electrostatic}}$ and acoustic power absorbed by the perfectly-matched layer $P_{\mathrm{acoustic}}$ are calculated. The quality factor set by piezoelectric-loss is then given by
\begin{equation}
    Q = \frac{\omega E_{\mathrm{electrostatic}}}{P_{\mathrm{acoustic}}}.
\end{equation}

The two-dimensional nature of the simulation means that acoustic modes with out-of-plane (i.e. along the waveguide) propagation or strain are neglected. Modes with an out-of-plane propagation direction couple weakly to the microwave resonator because the capacitor is much longer than the acoustic wavelength at the relevant ${\sim} \mathrm{GHz}$ frequencies. Modes with out-of-plane stress also couple weakly to the microwave resonator for X-cut lithium niobate because of lithium niobate's piezoelectric coefficients. For example, when applying the electric field along the Z crystal axis in our device the $d_{33}$ piezoelectric coefficient, which creates in-plane stress, dominates over other components.

\section{Measurement setup and transduction efficiency calibration}

\begin{figure*}
\centering
\includegraphics[width=\linewidth]{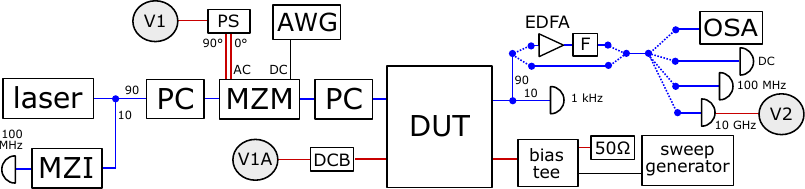}
\caption{Simplified diagram of the transduction measurement setup. Laser emission passes through polarization controllers (PC) and is modulated by a Mach-Zehnder modulator (MZM). A fraction of the emission is sent to a 100 MHz-bandwidth detector after passing through a Mach-Zehnder interferometer (MZI) of ${\sim}1$ m path length difference to precisely track sweeps of laser wavelength. The MZM is arranged for either GHz-frequency optical single-sideband modulation using a phase-shifted (PS) dual drive through a high-frequency port, or low-frequency amplitude modulation through a bias port, controlled by an arbitrary waveform generator (AWG). Focusing grating couplers (${\approx} 10\, \mathrm{dB}$ insertion loss) couple light from optical fibers into the device under test (DUT), which is cooled to $T\approx 1\,\mathrm{K}$ inside a closed-cycle cryostat. The light collected from the DUT is split into an analysis arm ($90\%$) and a $1 \ \mathrm{kHz}$ photoreceiver ($10\%$), whose signal is used to lock the laser frequency to an optical mode. The analysis arm passes through several optical switches (dotted blue lines) which allow for optional and repeatable insertion of an erbium-doped fiber amplifier (EDFA) and optical filter (F, 0.2nm bandwidth). The analysis arm can be sent to an optical spectrum analyzer (OSA) for sideband calibration, a DC optical power meter for power calibration, a $100\ \mathrm{MHz}$ photoreceiver for measuring transmission spectra, or a $10 \ \mathrm{GHz}$ photoreceiver for detecting transduction. The bias voltage of the DUT is controlled by a sweep generator through a bias tee. A vector network analyzer (VNA) can be connected to microwave port V1A, which is protected by a DC block (DCB) capacitor, to excite the DUT. The upconverted optical signal can be detected at port V2. In an alternative measurement setup, the transmission of an optical sideband can be monitored by connecting the VNA to the optical single-sideband modulator (port V1).} 
\label{fig:setup}
\end{figure*}

Details of the measurement setup are shown in Figure \ref{fig:setup}. To calibrate the transduction efficiency, we perform the following procedure before every set of measurements. 

First, with the laser frequency detuned far from the optical resonance, we measure the optical power into and out of the DUT. After correcting for measured asymmetric losses in the optical fibers going into the cryostat, we assume the loss at both input and output grating couplers to be symmetric. Based on measurements of a large number of grating couplers, we estimate the coupler-to-coupler variation in insertion loss to be less than $0.4\,\mathrm{dB}$. Next, we measure the optical power arriving at the output of the analysis arm using the DC power meter. These measurements allow us to estimate the optical insertion loss from the DUT to the end of the analysis arm $\eta_{\mathrm{optical}} = \eta_{\mathrm{coupler}} \cdot \eta_{\mathrm{fiber}}$, as well as the on-chip optical power.

Next, we calibrate the response of the $10 \, \mathrm{GHz}$-bandwidth photoreceiver by using port V1 to generate a single optical sideband. We measure the signal in the analysis arm using the high-resolution optical spectrum analyzer (which allows us to directly measure the relative power of the sideband and carrier $P_{\mathrm{sideband}}/P_{\mathrm{pump}}$), the calibrated DC power meter (which measures the total power $P_{\mathrm{sideband}}+P_{\mathrm{pump}}$), and the $10 \, \mathrm{GHz}$ photoreceiver. These measurements allow us to estimate the detector response parameter $A_{\mathrm{det.}}$, defined so that $ P_{\mathrm{det.}}= A_{\mathrm{det.}}P_{\mathrm{sideband}}P_{\mathrm{pump}}$.

During the transduction measurement, when the laser frequency is locked to an optical mode, measurements of the total optical power and the $10 \, \mathrm{GHz}$ photoreceiver response allow us to infer the power in the upconverted optical sideband $P_{\mathrm{sideband}}$, based on the above photoreceiver calibration. The gain provided by the erbium-doped fiber amplifier (EDFA), if in use, can be estimated by measuring the photoreceiver response with and without the EDFA in the optical path. Finally, the calibrated transduction efficiency is given by
\begin{equation}
    \eta = \frac{\omega_mP_{\mathrm{sideband}}}{\omega_oP_{\mathrm{in}}\eta_{\mathrm{optical}}\eta_{\mathrm{cable}}},
\end{equation}
where $P_{\mathrm{in}}$ is the input microwave power at port V1A and $\eta_{\mathrm{cable}}$ is the measured insertion loss from port V1A to the DUT.

\section{Microwave resonator nonlinearity}
\begin{figure}[htbp]
\centering
\includegraphics[width=\linewidth]{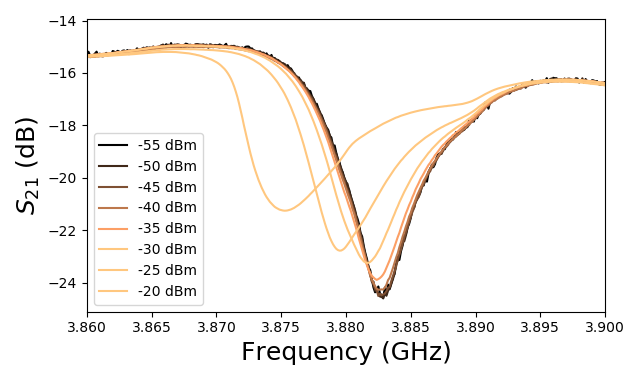}
\caption{Effect of microwave power on transmission spectrum. Microwave input powers above about $-40\  \mathrm{dBm}$ at port V1A ($\sim -48\ \mathrm{dBm}$ on-chip) produce distorted transmission spectra due to nonlinear dynamics. \cite{abdoNonlinear2006}} 
\label{fig:mwPower}
\end{figure}

The superconducting NbN film is deposited using DC magnetron sputtering at room temperature with an RF bias on the substrate holder. The film has a thickness of $\sim$44\,nm, room-temperature sheet resistance of 52\,$\Omega$/square, and a transition temperature $T_\mathrm{c}$ of $\sim$10\,K. At high microwave powers, the superconducting resonator undergoes nonlinear oscillations \cite{abdoNonlinear2006}, as shown in Fig.~\ref{fig:mwPower}. In the actual experiment, the drive power is kept below -30\,dBm, and the nonlinear dynamics are therefore small. 

\section{Interventions for improved transducer performance}

Table \ref{tab:interventions} lists several relatively straightforward interventions that can be made to improve the performance of our transducer. The predicted efficiency enhancement for each intervention in the table assumes the transducer operates in the low cooperativity limit.

\begin{table*}[htbp]
\centering
\caption{\bf Predicted improvements in device parameters}
\begin{tabular}{cccc}
\hline
Parameter & Improvement factor & Efficiency enhancement & Notes \\
\hline
$\kappa_{\mathrm{optical}}$ & $10$ & $10^2$ & Optical quality factor $Q=10^7$ \\
$\kappa_m$ & 10 & 10 & By suspending lithium niobate layer \\
Block scattered light & - & 10 & For $100$ \textmu W of on-chip pump power \\
Electrode coverage $\alpha$ & 2 & 4 & \\
Single-sided MW coupling & 2 & 2 & Improves effective MW loss \\
Microwave capacitance & 2 & 1.5 & Using a high-impedance inductor\\
Resonant optical pump & - & 1.5 & \\

\hline
Total & & $\sim 10^5$ &  \\
\hline
\end{tabular}
  \label{tab:interventions}
\end{table*}

\end{document}